\documentclass{ws-procs975x65}

\def\beq{\begin{equation}}
\def\eeq{\end{equation}}

\begin{document}

\title{Electric-magnetic duality in the quantum theory}

\author{Adrian del Rio$^*$}

\address{Centro de Astrof\'isica e Gravita{c}\~ao (CENTRA), Departamento de F\'isica, Instituto Superior T\'ecnico  (IST), Universidade de Lisboa - Lisboa 1049-001, Portugal\\
$^*$E-mail: adriandelrio@tecnico.ulisboa.pt}

\begin{abstract}
It is known that an electric-magnetic duality transformation  is a symmetry of the classical source-free Maxwell theory in generic spacetimes. This provides a conserved Noether charge, physically related to the polarization state of the electromagnetic field. We shall argue that this conservation law fails to hold at the quantum level in presence of a background classical gravitational field with non-trivial dynamics, as determined by the Chern-Pontryagin scalar. This is the spin 1 analog of the chiral anomaly for massless Dirac fermions.  
\end{abstract}


\bodymatter

\section{The classical electric-magnetic Noether symmetry}\label{aba:sec1}

It is known from elementary courses in electrodynamics that Maxwell equations without electric charges and currents, as well as the electromagnetic stress-energy tensor, are manifestly invariant under and an ``exchange" of the electric $\vec E$ and magnetic $\vec B$ fields, given by $\vec E \to \vec B$, $\vec B \to -\vec E$. In a manifestly covariant language, this amounts to say that 
\beq
\nabla_{\mu}F^{\mu\nu}=0, \hspace{1cm}  \nabla_{\mu}{^{\star}F}^{\mu\nu}=0, \label{MaxwellEq}
\eeq
together with $T_{\mu\nu}=\frac{1}{2}\left[F_{\mu\rho}F^{\rho}_{\hspace{0.15cm}\nu}+{^{\star}F}_{\mu\rho}{^{\star}F}^{\rho}_{\hspace{0.15cm}\nu}  \right]$, where $F$ denotes the electromagnetic  2-form field and ${^{\star}F}$ its Hodge-dual, remain invariant under the discrete transformation  $F_{\mu\nu} \to {^{\star}F}_{\mu\nu}$, $ {^{\star}F}_{\mu\nu} \to - F_{\mu\nu}$.  Because this discrete transformation is nothing but the Hodge-duality mapping between 2-forms in a 4-dimensional spacetime, this  invariance is usually regarded as the duality of Maxwell theory of electromagnetism.

Perhaps more interesting than this is that the  invariance actually extends to  {\it continuous} $SO(2)$ rotations between the fields, $F \to F\cos\theta +{^{\star}F}\sin \theta$, called electric-magnetic  transformations. This continuous transformation  not only leaves the above equations invariant, but it is also  a symmetry of the source-free Maxwell action functional
\beq
S[A]  = \int d^4x \sqrt{-g} F_{\mu\nu}F^{\mu\nu}\, , \label{MaxwellA}
\eeq
as first found by Calkin \cite{Calkin1965}. Here we introduced  the electromagnetic potential  $A$,   defined by $F=dA$.  Some years later, this was reanalyzed in greater detail by Deser and Teitelboim \cite{DeserTeitelboim1976} using the canonical approach, showing in particular that the symmetry holds even if the electromagnetic field propagates in a curved spacetime background.  Noether's theorem enters  the scene  providing a conserved charge and current, with valuable physical information. 

The Noether current takes the form (valid on-shell) \cite{AgullodelRioPepe2017a}
 \beq
j_D^{\mu} \approx \frac{1}{2}\left[A_{\nu}{^{\star}F}^{\mu\nu}-F^{\mu\nu}Z_{\nu}\right] \, ,
\eeq
(where $Z_{\nu}$ is defined by $E^{\mu}=-\epsilon^{\mu\alpha\beta}\nabla_{\alpha}Z_{\beta}$) and it is conserved on-shell: 
\beq
\nabla_{\mu}j^{\mu}_D=-Z_{\nu} \nabla_{\mu}F^{\mu\nu}\approx 0. \label{divergence}
\eeq 
The electric-magnetic symmetry of the theory is generated by a conserved charge $Q_D$,  which is the integral on a Cauchy surface of the ``zero'' component of the previous current:
\beq
Q_D =-\frac{1}{2} \int_{\Sigma_t} d\Sigma_t (A_{\mu} B^{\mu}-E^{\mu}Z_{\mu})  \, ,
\eeq
and satisfies $\delta H=\{H,Q_D\}\approx 0$.
This charge is gauge-invariant, it is a constant of motion $\dot Q\approx 0$, and  it is not difficult to check that generates the correct  transformation rule for the electric and magnetic fields in phase space:
\beq
\delta E=\{E,Q_D\}=B\, , \hspace{1cm}  \delta B=\{B,Q_D\}=-E \, .
\eeq

In Minkowski spacetime, this charge shows interesting physics. Expanding the fields in Fourier modes with positive $h_+(k)$ and negative $h_-(k)$ helicity amplitudes, the equation above reduces to
\beq
Q_D= \int_{\mathbb R^3} \frac{d^3k}{(2\pi)^3\, k}  \left[ |h_+(\vec k)|^2-|h_-(\vec k)|^2\right] \, .
\eeq
As we can see, the Noether charge accounts for the net difference between right- and left- handed circularly polarized radiation (or photons, in the quantum theory).  This is recognized as the usual V-stokes parameter, describing the polarization state of electromagnetic radiation. In some  contexts this is also known as optical helicity \cite{BarnettCameronYao2012}.  One concludes that, as a direct consequence of the classical electric-magnetic duality symmetry, this physical magnitude is a constant of motion of source-free electrodynamics.

\section{The anomaly in the quantum theory}

An important question now is whether the symmetry still holds in the quantum theory, i.e. if $\left< \nabla_{\mu}j^{\mu}_D\right>=0$ for any given quantum vacuum state. Failing to satisfy this will imply that the electric-magnetic transformation is not a quantum symmetry. Given the well-known and diverse examples of anomalous symmetries in quantum field theory, this is not a trivial issue.

\subsection{Why is an  anomaly expected?} \label{aba:sec2}

Anomalies in quantum field theory were first discovered by Adler, Bell, and Jackiw\cite{ABJ} for the Dirac chiral symmetry in the context of spinor quantum electrodynamics, and immediately after by Kimura\cite{Kimura} in the gravitational framework. The underlying reason for their appearance is renormalization. In short,  the calculation of expectation values of quadratic field operators runs into problems due to UV divergences. They are ill-defined and as such we need renormalization to extract a well-defined physical quantity from them. The Noether currents are a prototype example of that. The key point is that renormalization subtractions do not necessarily respect the classical equations of motion. Then, quantum fluctuations lead to non-trivial off-shell contributions to the divergence of the current (\ref{divergence}),  causing the breakup of the classical invariance. 

Of course, not every symmetry of a field theory is ruined by quantum fluctuations. One of the key points that makes the appearance of an anomaly very likely in this situation is the fact that the electric-magnetic transformation resembles an ordinary chiral transformation when working with self- and antiself- dual field variables. Indeed, if we define the self- dual fields $\vec H_{\pm}=\frac{1}{\sqrt{2}}\left[ \vec E\pm i \vec B\right]$ then
\beq
\vec H_{\pm} \to e^{\mp i \theta} \vec H_{\pm}\, ,
\eeq
is an electric-magnetic rotation of the fields. Furthermore, under a Lorentz transformation both fields decouple, and their transformation rules are related to the two irreducible representations of the Lorentz group for fields of spin $1$, written in standard terminology as $(0,1)$ and $(1,0)$.
Compared with massless fermions of spin $1/2$, $\vec H_+$ is the analog of a right-handed Weyl spinor ---that transforms under the $(0, 1/2)$ Lorentz representation--- and $\vec H_-$ is the analog of a left-handed Weyl spinor. Thus, there is, indeed, full analogy with  the fermion chiral transformation. Since this classical symmetry is known to be anomalous in the quantum theory if the Dirac field propagates in a curved spacetime,  it becomes plausible that a similar feature will happen in the electromagnetic case too.

\subsection{Derivation of the quantum anomaly} \label{aba:sec3}

 The strategy that we follow to compute $\left< \nabla_{\mu}j^{\mu}_D\right>$  is to mimic as close as possible the  chiral anomaly for Dirac fields.

The previous arguments strongly suggest to use self- and antiself- dual variables. Their use  actually makes the dynamics of the theory significantly more transparent. Source-free Maxwell equations (\ref{MaxwellEq}) can be rewritten as
\beq
\alpha^{\mu\nu}_I \nabla_{\mu}H_+^{I}=0 \, ,
\eeq
with $\alpha^{\mu\nu}_I$ some matrices satisfying the $su(2)$ Lie algebra (again, this resembles the spin 1/2 theory of Weyl spinors). As in the standard formulation, this equation can be solved by introducing potentials.  Defining complex potentials by $H^I_+=i \epsilon^{I\mu\nu}\nabla_{\mu}A_{+ \, \nu}$ the equations of motion read now \cite{AgullodelRioPepe2017a}
\beq
\bar \alpha^{\mu\nu}_{\dot I} \nabla_{\mu}A^+_{\nu}=0 \, .
\eeq

The analogy with the Dirac theory goes even further. The standard Maxwell action functional (\ref{MaxwellA})   can be rewritten in terms of complex variables in the following form \cite{AgullodelRioPepe2017a}
\beq
   S[A^+,A^-]=-\frac{1}{4}\int d^4x\sqrt{-g}\ \bar \Psi \, i\beta^{\mu}\nabla_{\mu}\Psi \label{DiracS} \, ,
\eeq
where 
\beq \label{Psi}
\Psi=\left( {\begin{array}{c}
 A^+ \\ H_+  \\ A^-  \\ H_- \\
  \end{array} } \right)   \, , \hspace{.5cm} \bar \Psi =   ( A^+,   H_+,    A^-,   H_- ) \, , \hspace{.5cm} \beta^{\mu}= i\,     \left( {\begin{array}{cccc}
 0 & 0 & 0  & \bar\alpha^{\mu}  \\
  0  & 0 &  -\alpha^{\mu}  &0 \\
 0 &  \alpha^{\mu}  & 0  & 0 \\
- \bar  \alpha^{\mu}  & 0  & 0  &0 \\
  \end{array} } \right)  
\ . \eeq
These new $\beta^{\mu}$ matrices satisfy the familiar Clifford algebra $\{\beta^{\mu},\beta^{\nu}\}=2g^{\mu\nu}$, and allow us to define a chiral matrix  $\beta_5=\frac{i}{4!}\epsilon_{\mu\nu\alpha\rho}\beta^{\mu}\beta^{\nu}\beta^{\alpha}\beta^{\rho}$. Then the chiral-type transformation of the fields
\beq
\Psi \to e^{i\beta_5 \theta}\Psi \, ,
\eeq
defines the electric-magnetic rotation, and it is a symmetry of the above action. By applying Noether's theorem we immediately get the current $j^{\mu}_D=-\frac{1}{4}\bar \Psi \beta^{\mu}\beta_5 \Psi$, which agrees with the one presented above.

Now we are equipped with sufficient tools to calculate the vacuum expectation value of (\ref{divergence}). We  took  two different routes to obtain the result, each one illustrating different aspects of the anomaly. 

The first method is a direct computation:  we identify and regularize UV divergences, and  subtract them in a covariant way. This is achieved by writing the divergence of $j_D^{\mu}$ in terms of the Wightman two point function $S(x,x')=\left< \bar \Psi(x) \Psi(x')\right>$, then subtract the corresponding UV asymptotic expansion up to 4th order in derivatives of metric (which contain UV divergences), and then taking the limit in which the two points merge:
\beq
\left< \nabla_{\mu} j^{\mu}_D \right>_{ren}  =  \left< \nabla_{\mu}j^{\mu}_D\right> - \left< \nabla_{\mu}j^{\mu}_D\right>_{(4)}  \nonumber\\
= \lim_{\vec x\to \vec x'  \lambda \to 0 } \frac{i}{2} \lambda {\rm Tr} \beta_5 \left( S(x,x',\lambda) - S(x,x',\lambda)_{(4)} \right) \, .
\eeq
 The asymptotic expansion is evaluated using the Heat Kernel method \cite{parker-toms}.
 After taking limits a non-vanishing term survives, yielding an anomalous non-conservation of the duality current \cite{AgullodelRioPepe2017a}:
 \beq
 \left< \nabla_{\mu} j^{\mu}_D \right>_{ren}  = -\frac{\hbar}{96\pi^2}R_{\mu\nu\rho\sigma}{^{\star}R}^{\mu\nu\rho\sigma}  \label{anomaly} \, .
 \eeq
This non-vanishing result arises entirely from the subtraction terms in the renormalization process, which are independent of the vacuum state.

The second approach relies on Fujikawa's path integral viewpoint\cite{Fujikawa, Fbook}. The partition function of the theory,
\beq
Z=\int d\mu[\Psi,\bar \Psi] e^{i S[\Psi, \bar \Psi]} \, ,
\eeq
 remains invariant under a change of variables in phase space. An electric-magnetic duality rotation is a canonical transformation in phase space and as such it must leave $Z$ invariant.
Noether's theorem states that the action changes as
\beq
S[\Psi',\bar \Psi']=S[\Psi,\bar \Psi]-\int d^4x\sqrt{-g}\theta(x) \left< \nabla_{\mu} j_D^{\mu}\right>_{ren} \, .
\eeq
If everything else under the integral remained invariant, one would conclude that the current is conserved.  However, as first pointed out by Fujikawa, the measure of the integral is not necessarily invariant and could change by a non trivial Jacobian,   $d\mu[\Psi',\bar \Psi']=J d\mu[\Psi,\bar \Psi]$, $J\neq 1$. This is what yields  the anomalous non conservation of the current in this framework. A detailed analysis following this approach gives back  equation (\ref{anomaly}) again.

\section{Conclusions and final comments} \label{aba:sec5}

We showed that quantum fluctuations of the electromagnetic field spoil the classical conservation law of the Noether current associated to electric-magnetic rotations due to spacetime curvature, as calculated in (\ref{anomaly}). {The result points out that  $Q_D$ is no longer a constant of motion in the quantum theory: the gravitational dynamics should distinguish between right- and left-handed photons.} Physical implications of this quantum effect are still under investigation, but the analogy with the chiral anomaly \cite{Blaeretal1981} suggests that this is related to spontaneous  electromagnetic circular polarization, as a result of asymmetric creation of right/left photons  from the quantum vacuum due to background dynamics. If confirmed, our next goal will be to study the stimulated contribution to this asymmetric particle creation. This could be of  interest in different astrophysical scenarios\cite{AgullodelRioPepe2017b}, ranging from stellar gravitational collapse to mergers of compact objects.

\section*{Acknowledgments}

The author is  grateful to I. Agullo and J. Navarro-Salas for useful comments on the manuscript, and acknowledges financial support provided under the ERC Consolidator Grant ``Matter and strong-field gravity: New frontiers in Einstein's theory", no. MaGRaTh-$646597$.

\end{document}